# Exploring a Physics-Informed Decision Transformer for Distribution System Restoration: Methodology and Performance Analysis

Hong Zhao, Jin Wei-Kocsis, *Member, IEEE*, Adel Heidari Akhijahani, and Karen L Butler-Purry, *Fellow, IEEE*

*Abstract*— Driven by advancements in sensing and computing, deep reinforcement learning (DRL)-based methods have demonstrated significant potential in effectively tackling distribution system restoration (DSR) challenges under uncertain operational scenarios. However, the data-intensive nature of DRL poses obstacles in achieving satisfactory DSR solutions for large-scale, complex distribution systems. Inspired by the transformative impact of emerging foundation models, including large language models (LLMs), across various domains, this paper explores an innovative approach harnessing LLMs' powerful computing capabilities to address scalability challenges inherent in conventional DRL methods for solving DSR. To our knowledge, this study represents the first exploration of foundation models, including LLMs, in revolutionizing conventional DRL applications in power system operations. Our contributions are twofold: 1) introducing a novel LLM-powered Physics-Informed Decision Transformer (PIDT) framework that leverages LLMs to transform conventional DRL methods for DSR operations, and 2) conducting comparative studies to assess the performance of the proposed LLM-powered PIDT framework at its initial development stage for solving DSR problems. While our primary focus in this paper is on DSR operations, the proposed PIDT framework can be generalized to optimize sequential decision-making across various power system operations.

## I. INTRODUCTION

IN recent years, reliance on a stable and continuous power supply reaching unprecedented levels across all sectors of industry and daily life. Consequently, a reliable and resilient power supply has become paramount for various critical infrastructures such as hospitals, transportation systems, communication networks, and manufacturing facilities. At the same time, the increasing frequency and severity of extreme events, such as extreme weather, natural disasters, and cyber-attacks, pose critical challenges for maintaining the resilience and reliability of modern power systems. For instance, in February 2021, severe winter storms in Texas caused a massive electricity system outage, resulting in at least 57 deaths and over $195 billion in property damage, according to a report by the University of Texas at Austin [1]. These challenges highlight the urgent need to enhance the resilience of power grids to ensure they can withstand and recover from severe disruptions. Given that the distribution system serves as a crucial link in the power delivery process from the transmission grid to end-users, research into resiliency in distribution systems is pivotal for enhancing grid resilience and mitigating the impacts of power outages.

One key indicator of resiliency in distribution systems is the ability to restore service to critical loads after disruptions on the main grid [2]. Following power outages, distribution system restoration (DSR) aims to rapidly restore affected loads by leveraging advanced emerging controls once the outages are isolated. Forming microgrids (MGs) with dynamic boundaries as a service restoration strategy is a promising solution for enabling effective DSR [3-5]. By incorporating various energy sources, distributed generators (DGs), along with remotely controlled switches, a distribution system can be partitioned into multiple self-sufficient MGs. This approach enhances the system's restoration capability and maintains power supply continuity to critical loads, thereby significantly improving overall grid resilience. In this paper, we focus on advancing the DSR solution through sequential formation of MGs with dynamic boundaries, for enhancing resiliency of distribution systems.

The existing methods for solving DSR problems can be generally group into four categories: 1) mathematical programming methods, such as mixed-integer programming (MIP)-based methods. These methods formulate the DSR problems as mixed-integer linear or non-linear programming problems and solve them using off-the-shelf solvers [6-9,28]. While these methods ensure optimality of DSR solutions, they typically require accurate physical models of the distribution systems, which are not always available in dynamic and uncertain operation scenarios. Additionally, the computational complexity can increase dramatically with the number of controllable components, which limit the scalability of the solutions; 2) heuristic methods. These methods leverage algorithms, such as genetic algorithms, Tabu search, and greedy algorithms, to search for satisfactory solutions [10-12]. Compared to mathematical programming, heuristic methods are capable of handling dynamic and uncertain operation environment. However, these methods also face the scalability issues for large-scale distribution systems. Additionally, the quality of the solutions can be sensitive to the hyperparameters

Hong Zhao and Jin Wei-Kocsis are with Department of Computer and Information Technology, Purdue University, West Lafayette, IN 47907 USA (e-mail: {zhao1211, kocsis0}@purdue.edu). Adel Heidari Akhijahani and Karen L Butler-Purry are with Department of Electrical and Computer Engineering, Texas A&M University, College Station, TX 77843 USA (e-mail: {adelheidari, klbutler}@tamu.edu).

of heuristic algorithms; 3) expert systems. These methods use knowledge-based techniques, such as rule-based systems and fuzzy logic systems, to achieve DSR solutions [13-15]. These methods are effective in providing consists and quick decision makings based on encoded knowledge. However, capturing, encoding, and updating expert knowledge base can be challenging; and 4) machine learning methods, especially deep reinforce learning (DRL)-based methods. Emerging DRL technologies have recently gained attention for enabling more efficient, adaptive, and robust DSR solutions in uncertain operation scenarios. These methods formulate decision making for DSR under uncertainties as a Markov decision process (MDP) or a partially observable Markov decision process (POMDP) and solved them iteratively using data-driven DRL techniques such as deep Q-learning, advantage actor critic (A2C) algorithms, and proximal policy optimization (PPO) algorithms [16-19]. While DRL-based methods have shown great potential in efficiently addressing DSR in uncertain operation scenarios, their data-intensive nature poses challenges in achieving satisfactory DSR solutions for large-scale complex distribution systems. Various research efforts have been conducted to tackle the scalability issues, with one main trend being formulating of the distribution systems as multi-agent system and then developing multi-agent DRL methods for DSR [5, 20, 21]. While these methods have demonstrated efficiency in addressing DSR in large-scale distribution systems, the coordination between the agents can introduce high communication overhead and operation complexity. Additionally, as indicated in [22, 23], achieving stable learning and convergence can be challenging due to the non-stationary nature of the multi-agent system.

Recently, we have observed the advances of emerging foundation models, such as large language models (LLMs), in revolutionizing various application domains, including virtual assistants, healthcare, and education [24-26]. In these applications, foundation models like LLMs have demonstrated their advanced capabilities, such as context awareness on long-term dependencies, generative sequence modeling, and large-scale high-dimensional data processing. Inspired by the immense potential of these models, we pose a research question (RQ): *Can we leverage the powerful computing capabilities of foundation models, especially LLMs, to address the previously discussed scalability issue of DRLs for solving DSR?* The work presented in [27] implies potential direction on addressing this RQ. In [27], an LLM-powered concept, Decision Transformer, was proposed to first transform conventional DRL by modeling it as a conditional sequence modeling problem, and then leverage a causally masked transformer originally developed for LLMs, such as GPT-x model, to generate optimal actions for the DRL by conditioning on desired returns, past states, and actions. Given the powerful computing capabilities of LLMs in context awareness for long-term dependencies, generative sequence modeling, and large-scale high-dimensional data processing, it is reasonable to investigate the concept of a Decision Transformer for addressing our RQ and explore a transformative computing solution for solving the DSR problem. However, to the best of our knowledge, no existing work has explored the capability of the Decision Transformer in any power system operations, including DSR. This gap may be due to the complex inherent physical constraints present in power systems.

To address this gap, in this paper, we aim to develop a novel Physics-Informed Decision Transformer (PIDT) for solving the DSR problem. As far as we know, this is the first paper that explores the capability of foundation models, including LLMs, in revolutionizing conventional DRLs for power system operations. While our focus is on the DSR problem, the proposed PIDT framework can be generalized to optimize sequential decision-making for other power system operations. The main contributions of our proposed work are twofold:

1) A novel PIDT framework is proposed as the first-ever effort on exploring the powerful computing capabilities of LLMs in transforming conventional DRL for DSR operations.
2) Comparative studies are conducted to analyze the performance of the proposed LLM-powered PIDT framework in its initial development stage for solving the DSR problem.

The next section illustrates the problem settings for our work. In Section III, we describe our proposed LLM-powered PIDT framework for the DSR problem. Section IV shows the case studies and performance evaluations of the proposed PIDT framework. Conclusions are presented in Section V.

## II. PROBLEM SETTINGS

In our work, we formulate the DSR problem as an MDP within DRL framework. Table I shows the definitions of the parameters and variables that will be used in this paper.

### A. DSR Problem Modeling

In the initial stage of this research, the proposed Distribution System Restoration (DSR) method is modeled as a sequential decision-making process involving sequences of control actions on switches to restore loads to their normal operational states. These sequences of switching actions, referred to as energization paths, are each associated with a single active distributed generator (DG). Specifically, each energization path begins at the switch connected to the associated DG and aims to maximize load restoration by forming multiple microgrids, while ensuring compliance with all operational and physical constraints. Additionally, to reduce the space of control actions in the modeling, we adopt the concept of node cell as introduced in [28]. A node cell is defined as a set of nodes that are interconnected directly by non-switchable lines. Consequently, all the lines and loads within a node cell will be energized simultaneously. Furthermore, in our current work, the constraints include: 1) the voltage limits, 2) power flow constraints, 3) DGs' generation capacities, and 4) topological constraints including prevention of loop formation and ensuring no node cell is visited more than once.

### B. DSR Problem Formulation within a DRL Framework

We further formulate the sequential decision-making process for energization paths in the DSR problem model as an MDP

| | TABLE I |
|---|---|
| | DEFINITION OF VARIABLES AND PARAMETERS |
| $s_t, S$ | State at time $t$, the state space |
| $a_t, \mathcal{A}$ | Action at time $t$, the action space |
| $\mathcal{R}$ | Reward |
| $\mathcal{P}$ | Transition probability |
| $r_t$ | Reward at time $t$ |
| $\gamma$ | Discount factor |
| $T$ | Time horizon |
| $L$ | Set of all the loads |
| $N$ | Set of all nodes |
| $C$ | Set of all node cells |
| $x_{i,t}^L, x_{i,t}^N, x_{i,t}^C$ | Energization status of Load $i$ at time $t$, energization status of Node $i$ at time $t$, energization status of node cell $i$ at time $t$ |
| $P_{l,t}^L, P_{l,t}^C, P_{l,t}$ | Active power of Load $l$ at time $t$ when restored, accumulative active power of all loads in node cell $l$ at time $t$ when restored, nominal active power of Load $l$ at time $t$. |
| $H_{i,t}^L$ | Squared voltage magnitude of Load $i$ at time $t$ |
| $H_i^{\min}, H_i^{max}$ | Minimum and maximum squared nodal voltage of Load $i$ |
| $V_{l,t}^{L,p}$ | Voltage penalty function of Load $i$ at time $t$ |
| $\pi$ | Policy function |
| $\theta$ | Learnable parameter for the policy function (or the proposed PIDT) |

within the framework of DRL. The formulated MDP can be described as $MDP = \{S, \mathcal{A}, \mathcal{P}, \mathcal{R}, \gamma, T\}$, where:

1) State $s_t \in S$ is defined as the available observation vector of the overall distribution system at time $t$. The vector consisting of current load restoration, the voltages of individual loads, the status of the operable switches, and the status of energization trajectories.
2) Action $a_t \in \mathcal{A}$ is defined as the control actions applied to the operable switches, specifically selecting which switch to activate, in order to maximize the load restoration during DSR operations while ensuring operational and physical constraints.
3) Transition probability $\mathcal{P}: S \times \mathcal{A} \times S \to [0,1]$: Given state $s_t$ and action $a_t$ at time $t$, the distribution system transits to state $s_{t+1}$ at time step $t+1$ according to the transition probability $P(s_{t+1}|s_t, a_t)$.
4) Time horizon $T$ defines the sequence of decision-making steps.
5) Discount factor $\gamma \in [0,1]$ is to formulate the importance of future reward.
6) Reward $\mathcal{R}: S \times A \to \mathbb{R}$ evaluates the effectiveness of action $a_t$ taken in state $s_t$ in achieving the objective of our DSR problem, which is to maximize the load restoration in the time horizon $T$ while adhering to operational and physical constraints. Therefore, the reward function will be formulated as:

$$r_t = R_t(s_t, a_t) = R_t^A(s_t, a_t) + w_p \times R_t^V(s_t, a_t), \quad (1)$$

where $R_t^A(s_t, a_t)$ is the reward related to the total active power restoration at $t$, which is defined as:

$$R_t^A(s_t, a_t) = \left(\sum_{l \in L} x_{l,t}^L P_{l,t}^L\right) \times \Delta t \quad (2)$$

By incorporating the concept of node cell, Eq. (2) can be rewritten as:

$$R_t^A(s_t, a_t) = \left(\sum_{l \in J} x_{l,t}^C P_{l,t}^C\right) \times \Delta t \quad (3)$$

Additionally, $R_t^V(s_t, a_t)$ is a penalty term to penalize actions that violate the voltage constraints. It is defined as:

$$R_t^V(s_t, a_t) = -\left(\sum_{l \in L} x_{l,t}^L V_{l,t}^{L,p}\right) \times \Delta t, \quad (4)$$

where

$$V_{l,t}^{L,p} = \max(0, H_{l,t}^L - H_l^{\max}) + \max(0, H_l^{\min} - H_{l,t}^L), \quad (5)$$

$V_{l,t}^{L,p}$ represents the penalty terms associated with individual loads to ensure their voltage magnitude do not violate the constraints. These constraints are formulated using squared nodal voltage range $\left[H_i^{\min}, H_i^{max}\right]$. The weight term $w_p$ is set to ensure that the penalty is comparable to the active restored power term. Additionally, in our current initial stage of development, the topological constraints are hard-coded with a check function to automatically filter out violated actions.

Within the DRL framework, the objective of the DSR problem is to adaptively learn an optimal policy $\pi^*$ that maximizes the expected sum of rewards $\mathbb{E}[\sum_{t=1}^T r_t]$ over the trajectory $\tau = (s_1, a_1, r_1, s_2, a_2, r_2 \ldots, s_T, a_T, r_T)$.

III. PROPOSED LLM-POWERED PIDT FRAMEWORK FOR DSR DECISION MAKINGS

We continue to introduce our proposed innovative LLM-powered PIDT framework that transforms our formulated DRL framework by exploring the powerful computing capabilities of LLMs. Figure. 1 illustrates the overview structure our proposed PIDT framework. As shown in Fig. 1, our proposed PIDT framework mainly consists of: 1) an encoder that consists of linear layers followed by an embedding layer; 2) a GPT-based causal transformer with a causal self-attention mask; and 3) a decoder. The encoder is developed to take a trajectory of rewards, states, and actions as input, and process the trajectory to generate token embeddings that will be fed into the GPT-based causal transformer model for further processing. To achieve this, the trajectory input $\hat{\tau}_{-k:t}$ of length $T$ of the encoder is formulated $\hat{\tau}_{-k:t} = \{\hat{R}_K, s_K, a_K, \hat{R}_{K+1}, s_{K+1}, a_{K+1}, \ldots, \hat{R}_{t-1}, s_{t-1}, a_{t-1}, \hat{R}_t, s_t\}$ as illustrated in Fig. 1, which is different from the trajectory formulated in the original DRL framework that is described in previous section. To appropriately formulate the trajectory input, it is essential to define a physics-informed return-to-go reward $\hat{R}_t$ such that 1) the trajectory $\hat{\tau}_{-k:t}$ can be effectively processed by the GPT-based causal transformer, and 2) the objective and physical/operational constraints for DSR are accurately characterized. To achieve this, we formulate $\hat{R}_t$ by transforming the reward $r_t$ for the original DRL framework, which is defined in Eq. (1), by defining the desired target return as $\hat{R}^* = \sum_{t=1}^{T}(\sum_{l\in L} P_{l,t})$, where is $P_{l,t}$ is nominal active power of Load $i$ at time $t$. Based on $\hat{R}^*$ and $r_t$ defined in Eq. (1), we are able to represent $\hat{R}_t$ in training and inference procedures as follows:

1) Training Procedure:
$$\hat{R}_t = \sum_{i=t}^{T} r_i \tag{6}$$

2) Inference Procedure:
$$\begin{cases} \hat{R}_0 = \hat{R}_1 = \hat{R}^* \\ \hat{R}_t = \hat{R}^* - \sum_{i=1}^{t-1} r_i \end{cases} \tag{7}$$

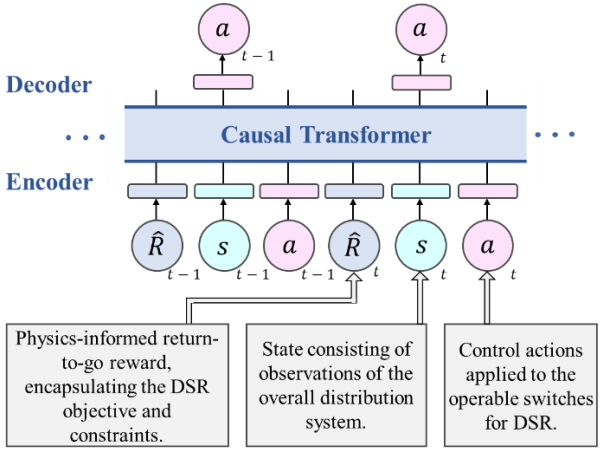

Fig. 1. Overview of the architecture of our proposed PIDT framework.

Where $t = 2, \ldots T$. After projecting the trajectory $\hat{\tau}_{-k:t}$ to the embedding dimension, the encoder forwards the token embeddings for $\hat{\tau}_{-k:t}$ to GPT-based casual transformer that is designed to generate a deterministic action at time $t$ such that $a_t = \hat{\pi}(\hat{\tau}_{-k:t})$, as shown in Fig. 1. The policy $\hat{\pi}$ within our PIDT framework is parameterized by the GPT-based casual transformer where the action sequences are generated via autoregressive modeling. The policy $\hat{\pi}$ is trained by minimizing the cross-entropy loss between the predicted actions and the ground-truth actions in a sampled batch of trajectory data. The output of the GPT-based casual transformer is fed to the decoder that projects the token embeddings of the predicted trajectory back to the original action space, resulting in the final predicted control actions on the operational switches for DSR.

**Model Training**. The overall training procedure is described as follows:

1) We prepare a dataset $D$ consisting of "offline" trajectories for DSR operations on the targeted distribution system. These trajectories can be collected from experts in power system domain or can be collected from simple off-line random walks to generate the sequences of control actions applied to the operable switches. These trajectories do not need to be optimal.

2) The minibatches of sequence length $K$ from the dataset $D$ will be fed to the PIDT framework to make decision on switching actions, and parameters of the PIDT framework will be updated according to the cross-entropy loss.

3) Repeat $M$ episodes for Step 2.

The procedure is shown in Algorithm 1.

| **Algorithm 1**: The PIDT Model Training |
|---|
| **Initialize**: Dataset $D$, PIDT model with learnable parameter $\theta$, sample size $L$, maximum number of episodes $M$, minibatch size $b$. |
| 1:    **for** episode $m = 1$ to $M$: |
| 2:       Sample a random minibatch of $b$ sequences of length $K$ from $D$. |
| 3:       Obtain predictions using $PIDT$ from batch $B$. |
| 4:       Calculate the cross-entropy loss between predicted and ground-truth values of action in sequence. |
| 5:       Update $\theta$. |
| 6:    **end for** |

**Model Inference**. The overall inference procedure is shown in Algorithm 2. It is worth noting that we always keep only the last $K$ time steps in the trajectory.

| **Algorithm 2**: Inference of the Trained PIDT Model |
|---|
| **Input**: PIDT model with trained parameter $\theta$, return-to-go $\hat{R}_0$, initial state of the distribution system $s_0$ |
| 1:    Set $\hat{R}_1 \leftarrow \hat{R}_0$, $s_1 \leftarrow s_0$, $\tau \leftarrow (s_1, \hat{R}_1)$ |
| 2:    **for** time step $t = 1$ to $T - 1$: |
| 3:       Obtain $a_t \leftarrow PIDT(\tau)$ and $r_t \leftarrow R_t(s_t, a_t)$ |
| 4:       Observe the next state $s_{t+1}$ after taking switching action $a_t$, and obtain $\hat{R}_{t+1} \leftarrow \hat{R}_t - r_t$. |
| 5:       Append $(a_t, s_{t+1}, \hat{R}_{t+1})$ to $\tau$, and keep the last $K$ time steps of $\tau$ (i.e., $\tau \leftarrow \tau_{-K:t+1}$) |
| 6:    **end for** |

## IV. PERFORMANCE EVALUATIONS

In this section, we evaluate the performance of our proposed PIDT framework in two case studies with the modified IEEE 13-bus system [29] and the modified IEEE 123-bus system [29], respectively. To evaluate the performance of the PIDT framework, we compare it with two benchmark DRLs, the PPO algorithm [30], and the A2C algorithm [31], in the DSR operation.

### A. Modified IEEE 13-Bus Test System

In this case study, we use a modified IEEE 13-bus test system that is available in the Open Distribution Simulator Software (OpenDSS). The system topology and its corresponding node-cell graph are shown in Fig. 2. As illustrated in Fig. 2, in this system, there is a substation located in a "sourcebus" that is connected to Bus Node 650, with an additional bus node (namely Bus Node 670) that connects between Bus Node 632 and 670. After reformed into node cells, there will be a total of node cells to be energized.

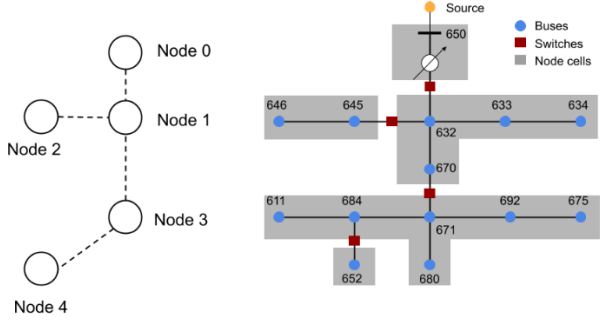

Fig. 2. Graph representation and physical topology of the modified IEEE 13-node test system.

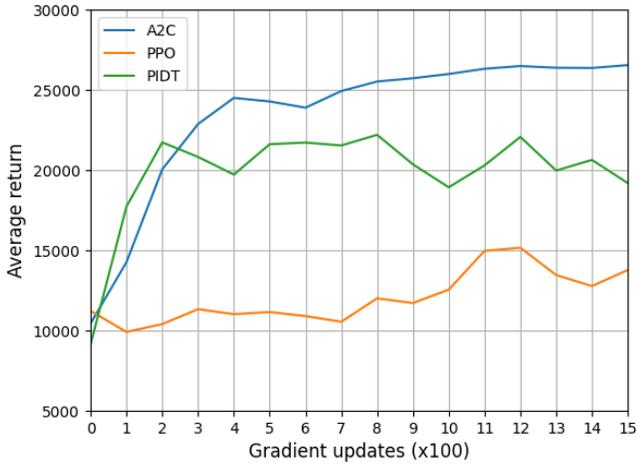

Fig. 3. Learning curves for the average returns of the first 1500 gradient updates using our PIDT method and the two other benchmark DRL methods, PPO and A2C, for the DSR operations in the modified 13-bus test system. The curve is updated per hundred gradient updates.

Additionally, based on objective and physical/operational constraints for the DSR operation, the ground-truth energization path in this case is determined as (Node 0) → (Node 1) → (Node 3) → (Node 4). And the corresponding final restored power should be 3006.509kW, which is considered as the objective of the DSR operation and is also the desired target return in our PIDT method. The learning curves for the average return of the first 1500 gradient updates by using our PIDT method as well as the other two benchmark DRL methods are shown are shown in Fig. 3. Further details about the evaluation results of the three methods are shown in Table II. Additionally, a bar char in Fig. 4 shows more insights of the simulation results, which presents the distribution of power restoration levels in the 50 independent trials using the three methods, respectively.

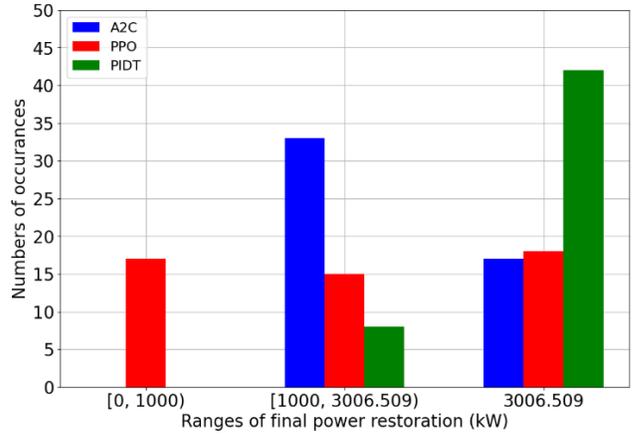

Fig. 4. Distribution of power restoration levels in the 50 independent trials using the three methods, respectively.

Example trajectories depicting optimal solutions and suboptimal solutions are shown in Appendix A.

TABLE II
FURTHER PERFORMANCE COMPARISON BETWEEN OUR PIDT METHOD AND OTHER TWO BENCHMARK DRL METHODS FOR THE DSR OPERATION IN THE MODIFIED IEEE 13-NODE TEST SYSTEM

| Methods | A2C | PPO | Our Method |
|---|---|---|---|
| Mean evaluation return | 26659.982 | 14810.724 | 20361.480 |
| Standard deviation of evaluation return | 643.940 | 8622.008 | 6130.914 |
| Optimal solution occurrence | 17 | 18 | 42 |
| Suboptimal solution occurrence | 33 | 32 | 8 |

From Figs. 3 and 4 and Table II, we observe that during the learning process, the A2C method demonstrates the quickest convergence and achieves the highest training return. However, it typically settles for a suboptimal solution, maintaining power restoration within the range of [1000, 3006] kW for the majority of trials. In contrast, our method, while not showing the fastest convergence or the highest training return, achieves optimal solutions in 42 out of 50 independent trials. This is reasonable since the return-to-go reward defined in our proposed PIDT method is different from the rewards in conventional DRL. Additionally, from the simulation results, we can also see that the performance of the PPO method falls between that of A2C and our PIDT methods.

*B. Modified IEEE 123-Bus Test System*

In this case study, we evaluate the performance of our PIDT method by comparing with the PPO- and A2C-based benchmark DRL methods in DSR operations for a modified 123-bus test system. The system topology and its corresponding node cell graph are shown in Fig. 5. As shown in Fig. 5, in the modified 123-bus test system, there are five energy sources in the system, two of which are substations located in Bus Nodes 150 and 350, and the three other sources are distributed generators (DGs) located in Bus Nodes 95, 250 and 450. Additionally, the system has a total of 15 node cells to be energized. We also incorporate the modification on the energization ability to the DG in Bus 250, such that this DG can only fully energize loads in Node Cell 3 and Node Cell 2. The DG will be automatically shut down if it tries to energize more loads other than those in Node Cell 3 and Node Cell 2.

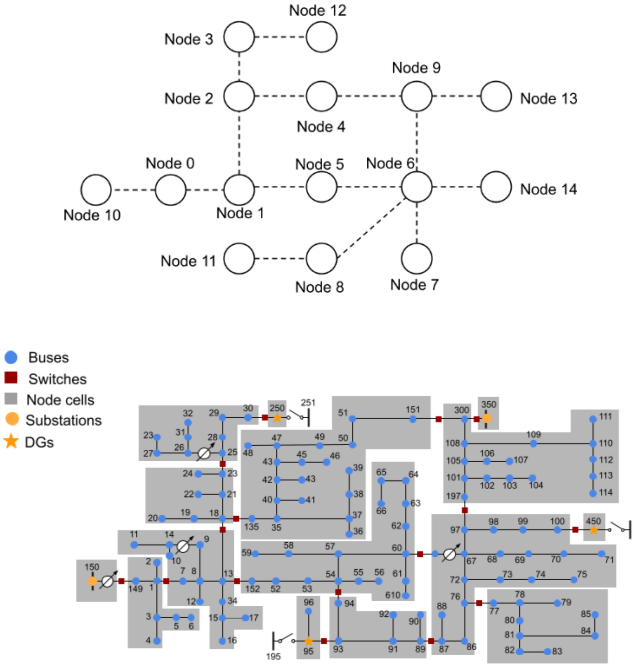

Fig. 5. Graph representation and physical topology of the modified IEEE 123-bus test system.

The actual total power of the system, if all the loads are powered properly, would be 3350.757 kW, which is considered as the objective of the DSR operation and is also the desired target return in our PIDT method. An expected optimal solution would power all the loads properly. However, due to the large0-scale and complex topology of the modified IEEE 123-bus test system, a simple optimal solution to power all the loads is very challenging. The learning curve of the first 6000 gradient updates using our PIDT method and the two benchmark DRL methods are shown in Fig. 6. As shown in Fig. 6, for the modified IEEE 123-bus system that has large-scale and complex topology, our method outperforms the other two methods in convergence rate and convergence reward in general. Further details about the evaluation results of the 3 methods are presented in Table III. As shown in Table III, our proposed PIDT method can achieve near-optimal solutions in 30 of the 50 independent trial, which outperforms the PPO and A2C methods (in this case study, the near-optimal solutions are defined as the solutions whose final restored power is more than 3000kW).

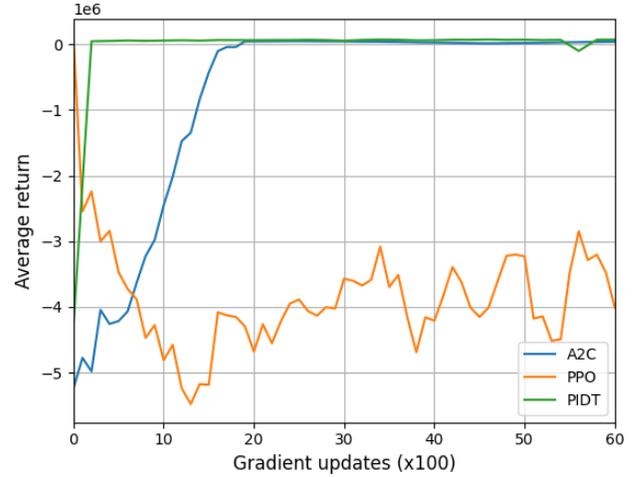

Fig. 6. Learning curve for the average return of the first 6000 gradient updates using our PIDT method and the two other benchmark DRL methods, PPO and A2C, for the DSR operations in the modified 123-bus test system. The curve is updated per hundred gradient updates.

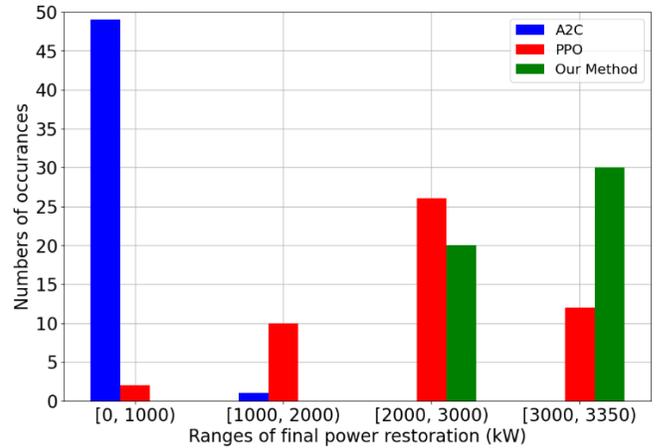

Fig. 7. Distribution of power restoration levels in the 50 independent trials using the three methods, respectively.

Furthermore, Fig. 7 provides deeper insights into the simulation results shown in Table III by illustrating the distribution of power restoration levels across 50 independent trials using the three methods. It reveals that our method achieves near-optimal power restoration in 30 trials and between 2000 kW to 3000 kW in the remaining 20 trials. In contrast, the solutions generated by the A2C-based method predominantly fall within the range of 0 kW to 1000 kW. The PPO-based method shows a more varied distribution across all three ranges, with 12 trials achieving near-optimal results. These findings underscore the superior performance of our method over the other two conventional DRL methods.

## V. CONCLUSIONS

While different DRL-based approaches have been developed for effectively addressing DSR challenges amidst uncertain operational conditions, the data-intensive nature of DRL presents barriers to achieving robust solutions for large-scale, complex distribution systems. This paper explores an innovative strategy inspired by the transformative impact of emerging foundation models, such as LLMs, across diverse domains. Specifically, we present a first-ever effort on exploring LLMs' powerful computing capabilities to tackle scalability challenges inherent in traditional DRL methods for DSR operations.

TABLE III
Evaluation results for the 3 methods on the modified IEEE 123-Node Test System

| Methods | A2C | PPO | Our Method |
|---|---|---|---|
| Mean evaluation return | 47414.036 | -4659896.17 | 100915.721 |
| Standard Deviation of evaluation return | 1675.582 | 5724583.67 | 19100.489 |
| Mean final restored power (kW) | 988.044 | 2452.702 | 2910.923 |
| Standard deviation of restored power (kW) | 29.400 | 662.671 | 369.444 |
| Near-optimal solution occurrence | 0 | 12 | 30 |

This study marks the pioneering application of foundation models, including LLMs, to revolutionize conventional DRL practices in power system operations. Our contributions are twofold: 1) introducing a novel PIDT framework that explore LLMs to transform conventional DRL methods for DSR operations, and 2) conducting comparative studies to analyze the performance of the LLM-powered PIDT framework in its initial development stage for solving DSR problem. Simulation results underscore the effectiveness of our proposed PIDT method in resolving DSR problems within large-scale distribution systems. In our ongoing work, we are leveraging the insights gained from our current initial-stage development to further enhance the scalability and resilience of our proposed PIDT method. While our primary focus in this paper lies in DSR, the proposed PIDT framework demonstrates potential applicability in optimizing sequential decision-making across a spectrum of power system operations.

APPENDIX

*A. Example solutions for the case study with modified IEEE 13-bus test system*

The step-by-Step illustrations on an optimal solution of the sequence of control actions generated by our method for the DSR operation in the modified 13-bus test system. are shown in Fig. 8, which also represents most of the solutions generated by our method.

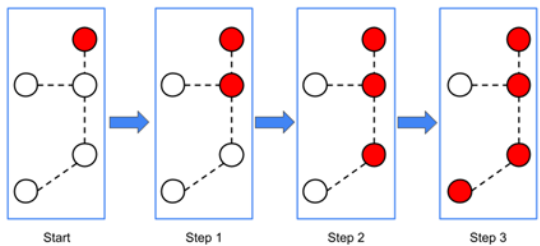

Fig. 8. Step-by-Step illustration on an optimal solution on the sequence of control actions generated by our method for the DSR operation in the modified 13-bus test system.

Additionally, Fig. 9 illustrates one representative suboptimal solution generated by our PIDT method. As part of our ongoing work, we are actively investigating these suboptimal solutions, which will give us valuable insights on improving our current initial-stage efforts.

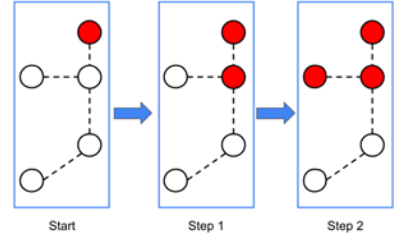

Fig. 9. Switching sequence of an example suboptimal solution generated by our method for the modified 13-bus test system.

*B. Example solutions for the case study with modified IEEE 123-bus test system*

The step-by-Step illustrations on a near-optimal solution of the sequence of control actions generated by our method for the DSR operation in the modified 123-bus test system. are shown in Fig.10. By following this switching sequence, all the loads in the system will be energized, achieving a final restored power of 3348.504kW. Additionally, Fig. 11 shows an example sub-optimal solution on switching sequence, generated from our method, for DSR operation in the modified IEEE 123-bus system. As shown in Fig. 11, the suboptimal solution cannot energize all the loads in the system. From Fig. 11, we can also observe that one reason that this solution fails to energize all the loads is that the energization paths formed from the control sequence cannot traverse all the node cells in the system. The final restored power of this example solution is 2843.730kW. In our ongoing work, we are exploring the strategy to improve the capability of our initial version of the PIDT method in addressing DSR problem in large-scale distribution system including the modified IEEE 123-bus system.

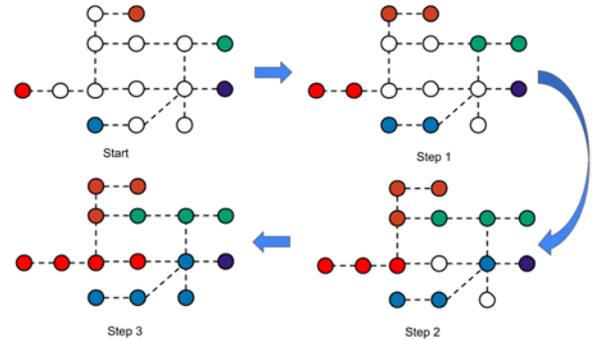

Fig. 10. Step-by-Step illustration on an example near-optimal solution on the switching sequence generated by our method for the DSR operation in the modified 123-bus test system.

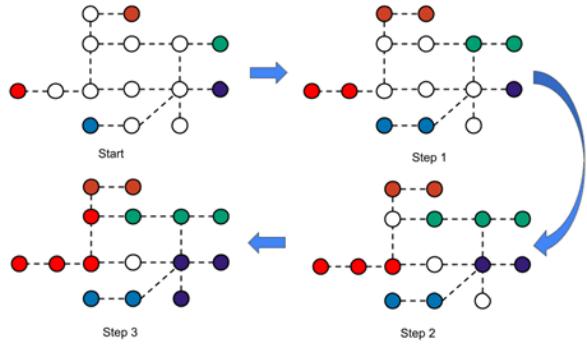

Fig. 11. Step-by-Step illustration on an example suboptimal solution on the switching sequence generated by our method for the DSR operation in the modified 123-bus test system.